\documentstyle[12pt,fleqn]{article}
\def\openone{\leavevmode\hbox{\small1\kern-3.8pt\normalsize1}}

\begin{document}

\begin{center}
{\LARGE Quantum State Disturbance vs.~Information Gain:\medskip

Uncertainty Relations for Quantum Information} \vfill

Christopher A. Fuchs$^*$\\[5mm]
{\it Center for Advanced Studies, Department of Physics and
Astronomy,\\
University of New Mexico, Albuquerque, NM 87131-1156}\\[5mm]
and\\[5mm]
Asher Peres$^\dagger$\\[5mm]
{\it Department of Physics, Technion---Israel Institute of
Technology, 32 000 Haifa, Israel}\\[7mm]
\end{center}\vfill

\noindent{\bf Abstract}\bigskip

When an observer wants to identify a quantum state, which is known to be
one of a given set of non-orthogonal states, the act of observation
causes a disturbance to that state. We investigate the tradeoff between
the information gain and that disturbance. This issue has important
applications in quantum crypto\-graphy. The optimal detection method,
for a given tolerated disturbance, is explicitly found in the case of
two equiprobable non-orthogonal pure states.

\vfill

\vfill

\noindent
$^*$\,Address in 1996: D\'epartement IRO, Universit\'e de Montr\'eal,
C.P. 6128, Succursale ``A'',\\ \hspace*{1mm} Montr\'eal, Qu\'ebec,
Canada H3C 3J7\\
$^\dagger$\,Electronic address: peres@photon.technion.ac.il\newpage

\begin{center}{\bf I. INTRODUCTION}\end{center}\bigskip

In the quantum folklore, the ``uncertainty principle'' is often taken
to assert that it is impossible to observe a property of a quantum
system without causing a disturbance to some other property. However,
when we seek the quantitative meaning of this vague declaration, all we
find are uncertainty {\it relations\/} such as
$\Delta x\,\Delta p\geq\hbar/2$,
whose meaning is totally different. Such a relation means that if we
prepare an ensemble of quantum systems in a well defined way (all in the
same way), and we then measure $x$ on some of these systems, and
independently measure $p$ on some {\it other\/} systems, the various
results obtained in these measurements have standard deviations, $\Delta
x$ and $\Delta p$, whose product is no less than $\hbar/2$. No reciprocal
``disturbance'' of any kind is involved here, since $x$ and $p$ are
measured on different systems (following identical preparations).

In this article, we shall give a quantitative meaning to the
heuristic claim that observation in quantum physics entails a necessary
disturbance. Consider a quantum system prepared in a definite way,
unknown to the observer who tests it. The question is how much
information the observer can extract from the system (how well he can
determine the preparation), and what is the cost of that information, in
terms of the disturbance caused to the system. This seemingly academic
question recently acquired practical importance, due to the development
of quantum cryptography [1--3], a new science which combines quantum
physics with cryptology. Following the established usage, the preparer
of the quantum state will be called Alice, the observer who wants to get
information while causing as little disturbance as possible will be Eve,
and a subsequent observer, who receives the quantum system disturbed by
Eve, will be called Bob. (In the cryptographical environment, Alice and
Bob are the legitimate users of a communication channel, and Eve is the
eavesdropper. The present paper discusses the situation in a general
way, from the point of view of what is possible in physics, and is not
concerned with any malicious motivations.)

First, we must define the notions of {\it information\/} and {\it
disturbance}. If Eve knows strictly nothing of $|\psi\rangle$ (the state
of the system that was prepared by Alice), she can gain very little
information by testing a single quantum system: for example, if she
chooses an orthonormal basis $|e_n\rangle$ and ``measures,'' in the von
Neumann sense of this term, an observable corresponding to that basis,
she forces the system into one of the states $|e_n\rangle$. In that
case, the answer only tells her that $|\psi\rangle$ before the
measurement was not orthogonal to the $|e_n\rangle$ that she found.
Meanwhile, the quantum state may be disturbed extensively in this
process. On the other hand, if Eve definitely knows that the initial
$|\psi\rangle$ is one of the orthonormal vectors $|e_n\rangle$, but she
does not know which one of them it is, she can unambiguously settle this
point by a {\it non-demolition measurement\/} [4], which leaves the
state of the system unchanged.

It is the intermediate case which is most interesting and has
applications to crypto\-graphy: Eve knows that Alice prepared one
of a finite set of states $|\psi_n\rangle$, with probability $p_n$.
However, these states are {\it not\/} all mutually orthogonal.
Before Eve tests anything, a measure of her ignorance is the Shannon
entropy $H=-\sum p_n\log p_n$.  She can reduce that entropy by suitably
testing the quantum system and making use of Bayes's rule for
interpreting the result (as explained in Section II).  The decrease in
Shannon entropy is called the {\it mutual information\/} that Eve has
acquired. The problem we want to investigate is the tradeoff between
Eve's gain of information, and the disturbance caused to the quantum
system.

A convenient measure for this disturbance is the probability that a
discrepancy would be detected by Bob, if he knew which state
$|\psi_n\rangle$ was sent by Alice, and tested whether the state that he
gets after Eve's intervention still is $|\psi_n\rangle$. In that case,
what Bob receives is not, in general, a pure state, but has to be
represented by a density matrix $\rho_n$. The disturbance (discrepancy
rate) detectable by Bob is

\begin{equation} D=1-\langle\psi_n|\rho_n|\psi_n\rangle.
\label{discrep}\end{equation}

Note that the mutual information and discrepancy rate, as defined above,
may not be the quantities that are most relevant to applications in
quantum cryptography [5]. An eavesdropper may not want to maximize
mutual information, but some other type of information, depending on the
methods for error correction and privacy amplification [6] that are used
by the legitimate users. Likewise, the protocol followed by Bob may not
be to measure $|\psi_n\rangle\langle\psi_n|$ for a particular $n$, but
to perform some other type of measurement.  In the present paper, we
have chosen mutual information and the discrepancy rate (1) for
definiteness (other possible choices are briefly discussed in the final
section).

In Section II of this article, we investigate the process outlined in
Fig.~1. Alice prepares a quantum system, in a state $\rho_{\rm A}$ (for
more generality, we may assume that this state is not pure and must be
represented by a density matrix). Eve likewise prepares a {\it probe\/},
with state $\rho_{\rm E}$. The two systems interact unitarily,

\begin{equation} \rho_{\rm A}\otimes\rho_{\rm E}\to\rho'
 =U\,(\rho_{\rm A}\otimes\rho_{\rm E})\,U^\dagger,
  \label{unitary} \end{equation}
and their states become entangled. Bob receives the system that Alice
sent, in a modified state,

\begin{equation} \rho_{\rm B}'={\rm Tr_E}(\rho'), \label{rB}
\end{equation}
where ${\rm Tr_E}$ means that the degrees of freedom of Eve's probe have
been traced out (since they are inaccessible to Bob). Bob may then test
whether this $\rho_{\rm B}'$ differs from the $\rho_{\rm A}$ that was
prepared by Alice. In the simple case where $\rho_{\rm A}$ is a pure
state, the discrepancy rate is $D=1-{\rm Tr} (\rho_{\rm A}\rho_{\rm
B}')$, as in Eq.~(\ref{discrep}).

How much information can Eve gain in that process? Her probe comes out
with a state

\begin{equation} \rho_{\rm E}'={\rm Tr_B}(\rho'), \label{rE}
\end{equation}
with notations similar to those in Eq.~(\ref{rB}). Now, to extract from
$\rho_{\rm E}'$ as much information as possible, Eve should not, in
general, perform a standard (von Neumann type) quantum measurement [7],
whose outcomes correspond to a set of orthogonal projection operators. A
more efficient method [8, 9] is to use a {\it positive operator valued
measure\/} (POVM), namely a set of non-negative (and therefore
Hermitian) operators $E_\mu$, which act in the Hilbert space of Eve's
probe, and sum up to the unit matrix:

\begin{equation} \sum_\mu E_\mu=\openone. \label{POVM} \end{equation}
Here, the index $\mu$ labels the various possible outcomes of the POVM
(their number may exceed the dimensionality of Hilbert space). The
probability of getting outcome $\mu$ is

\begin{equation} P_\mu={\rm Tr}(E_\mu\,\rho_{\rm E}'). \label{pmu}
\end{equation}
Such a POVM can sometimes supply more {\it mutual\/} information than a
von Neumann measurement.

Of course Eve cannot measure all the $E_\mu$ simultaneously, since in
general they do not commute. What she may do is to adjoin to her probe
an {\it ancilla\/} [8, 9] (namely an auxiliary system which does not
directly interact with the probe), and then to perform an ordinary von
Neumann measurement on the probe and the ancilla together (it is the
measuring apparatus that interacts with both of them). The advantage of
the POVM formalism, Eqs.~(\ref{POVM}) and (\ref{pmu}), is that it does
not require an explicit description of the ancilla (just as the von
Neumann formalism does not require an explicit description of the
measuring apparatus).

Here, the reader may wonder why we did not consider Eve's probe, and
her ancilla, and perhaps her measuring instrument too, as a single
object. The answer is that a division of the process into two steps has
definite advantages for optimizing it, as will be seen in detailed
calculations in Section III. Moreover, in some crypto\-graphical
protocols [1,~2], Alice must send to Bob, at a later stage, classical
information over a {\it public\/} channel. Eve, who also receives that
information, may in principle postpone the observation of her probe
until after that classical information arrives, in order to optimize the
POVM that she uses for analyzing her probe. This would not be possible
if the two steps in Fig.~1 were combined into a single one.\\[10mm]

\begin{center}{\bf II. INFORMATION--DISTURBANCE TRADEOFF}
\end{center}\bigskip

Let $\{|e_m\rangle\},\ m=1,\,\ldots\,N$, be an orthonormal basis for the
$N$-dimensional  Hilbert space of the system sent by Alice to Bob, and
let $\{|v_\alpha\rangle\}$ be an orthonormal basis for Eve's probe. The
dimensionality of the latter has to be optimized (see next section).
First, assume for simplicity that Alice sends one of the orthonormal
states $|e_m\rangle$, and that Eve's probe too is prepared in one of the
states $|v_\alpha\rangle$. (Results for other initially pure states can
be derived by taking linear combinations of the equations below. Mixed
states can then be dealt with by rewriting these equations in terms of
density matrices, and taking suitable weighted averages of the latter.)
The unitary evolution in Eq.~(\ref{unitary}) becomes, in the case we are
considering,

\begin{equation} |e_m,v_\alpha\rangle\to U\,|e_m,v_\alpha\rangle=
   \sum_{n\beta}A_{mn\alpha\beta}\,|e_n,v_\beta\rangle, \label{u0}
\end{equation}
where the notation

\begin{equation}
|e_m,v_\alpha\rangle\equiv|e_m\rangle\otimes|v_\alpha\rangle,
\end{equation}
was introduced, for brevity. The numerical coefficients
$A_{mn\alpha\beta}$ are the matrix elements of $U$:

\begin{equation} A_{mn\alpha\beta}=
 \langle e_n,v_\beta|U|e_m,v_\alpha\rangle. \label{U2A} \end{equation}
In the following, we shall drop the index $\alpha$: any mixed state
for Eve's probe can always be thought of as arising from a partial
trace over the degrees of freedom of a larger probe prepared in a pure
state.  We therefore assume that the probe's initial state is pure;
since the dimensionality of its Hilbert space still is a free variable,
this will cause no loss in generality.  Moreover, the final optimized
results are completely independent of the choice of that initial state
(because any pure state can be unitarily transformed into any other pure
state). The index $\alpha$ is therefore unnecessary. We thus obtain from
Eq.~(\ref{U2A}) the unitarity conditions

\begin{equation} \sum_{n\beta} A^*_{mn\beta}\,A_{m'n\beta}=\delta_{mm'}.
  \label{u1} \end{equation}

The final state, when Alice sends $|e_m\rangle$, can also be written as

\begin{equation} \sum_{n\beta}A_{mn\beta}\,|e_n,v_\beta\rangle=
   \sum_n |e_n\rangle\otimes|\Phi_{mn}\rangle, \end{equation}
where

\begin{equation} |\Phi_{mn}\rangle=\sum_\beta
 A_{mn\beta}\,|v_\beta\rangle,  \label{Phimu} \end{equation}
is a pure state of the probe. It is from these states and their linear
combinations that Eve will glean her information. Note that,
irrespective of the choice of $U$, i.e., for an arbitrary set of
$A_{mn\beta}$, there can be no more than $N^2$ linearly independent
vectors $|\Phi_{mn}\rangle$.  That is to say, the $N^2$ vectors
$|\Phi_{mn}\rangle$ span, at most, an $N^2$-dimensional space. Therefore
there is no point in using a probe with more than $N^2$ dimensions if
its initial state is taken to be pure.  (If the initial state of the
probe is a density matrix of rank $k$, the final states of that probe
span a Hilbert space of dimension not exceeding $kN^2$.) This point is
crucial for {\it any\/} optimization problem based solely on Eve's
measurement outcome statistics, not just the one for mutual information,
which is considered here.  It effectively delimits the difficulty of any
such problem, reducing it to computable proportions.

As we shall see, it is convenient to replace the $\beta$ index in
Eq.~(\ref{u1}), which may take $N^2$ values, by a pair of Latin indices,
such as $rs$, where $r$ and $s$ take the same $N$ values as $m$ or $n$.
We shall thus write $A_{mnrs}$ instead of $A_{mn\beta}$.

We now restrict our attention to the case where the quantum system
prepared by Alice is described by a two-dimensional Hilbert space (for
example, this may be the polarization degree of freedom of a photon).
The two dimensions will be labelled 0 and 1, so that $A_{mnrs}$ runs
from $A_{0000}$ to $A_{1111}$.  This quadruple index can then be
considered as a single binary number, and we thus introduce the new
notation:

\begin{equation} A_{mnrs}\to X_K,\qquad(K=0,\,\ldots\,,15). \end{equation}
The unitary relation (\ref{u1}) becomes
\begin{equation}\begin{array}{l}{\displaystyle
  \sum^7_{K=0}|X_K|^2=\sum^{15}_{K=8}|X_K|^2=1,}\medskip \\
  {\displaystyle
  \sum^7_{K=0} X^*_K\,X_{K+8}=0.}\end{array}\label{u2}\end{equation}

To further simplify the discussion, we assume that Alice prepares, with
equal probabilities, one of the pure states shown in Fig.~2(a):
\begin{equation}\begin{array}{l}
  |0\rangle=\cos\alpha\,|e_0\rangle+\sin\alpha\,|e_1\rangle,\medskip \\
  |1\rangle=\cos\alpha\,|e_1\rangle+\sin\alpha\,|e_0\rangle.\end{array}
 \label{signals} \end{equation}
By a suitable choice of phases, such a real representation can always be
given to any two pure states. Their scalar product will be denoted as

\begin{equation} S=\langle 0|1\rangle=\sin 2\alpha.
\label{S}\end{equation}

These notations are manifestly symmetric under an exchange of labels,
$0\leftrightarrow 1$. Since the two states are emitted with equal
probabilities, it is plausible that the optimal strategy for Eve is to
use instruments endowed with the same $0\leftrightarrow 1$ symmetry, so
that $\langle\Phi_{00}|\Phi_{01}\rangle
=\langle\Phi_{11}|\Phi_{10}\rangle$ and
$\langle\Phi_{00}|\Phi_{10}\rangle=\langle\Phi_{11}|\Phi_{01}\rangle$.
In particular, if Eve's $|v_{rs}\rangle$ basis is chosen in an
appropriate way (as explained below), the set of $A_{mnrs}$ also has the
01-symmetry, namely $A_{mnrs}=A_{\bar{m}\bar{n}\bar{r}\bar{s}}$, where
$\bar{m}=1-m$, etc. This relationship can be written as

\begin{equation} X_{15-K}=X_K, \end{equation}
and the unitary relations (\ref{u2}) become
\begin{equation}\begin{array}{l}{\displaystyle
  \sum^7_{K=0}|X_K|^2=1,}\medskip \\ {\displaystyle
  \sum^7_{K=0} X^*_K\,X_{7-K}=0.}\end{array}\label{u3}\end{equation}
Furthermore, we can safely drop the complex conjugation sign, since the
signal states (\ref{signals}) involve only real coefficients.  There is
no reason for introducing complex numbers in the present problem.

Still more simplification can be achieved by rotating the
$|v_\beta\rangle$ basis in a way that does not conflict with
01-symmetry. For example, in Eq.~(\ref{Phimu}), we may arrange that the
vectors $|v_{01}\rangle$ and $|v_{10}\rangle$ lie in the plane spanned
by the vectors $|\Phi_{01}\rangle$ and $|\Phi_{10}\rangle$, and that
they are oriented in such a way that
$\langle\Phi_{01}|v_{01}\rangle=\langle\Phi_{10}|v_{10}\rangle$, because
we want to have $A_{0101}=A_{1010}$ (no further rotation is then allowed
in that plane).  This is illustrated in Fig.~2(b). Note that we
automatically have
$\langle\Phi_{01}|v_{10}\rangle=\langle\Phi_{10}|v_{01}\rangle$, since
$|\Phi_{01}\rangle$ and $|\Phi_{10}\rangle$ have the same length, thanks
to the 01-symmetry. The vectors $|v_{00}\rangle$ and $|v_{11}\rangle$
are orthogonal to the plane spanned by $|v_{01}\rangle$ and
$|v_{10}\rangle$.  We likewise have to rotate them in their plane, so as
to have $\langle\Phi_{00}|v_{00}\rangle=\langle\Phi_{11}|v_{11}\rangle$,
and $\langle\Phi_{00}|v_{11}\rangle=\langle\Phi_{11}|v_{00}\rangle$.

With this choice of basis vectors for the probe, the $A_{mnrs}$
coefficients obey the 01-symmetry, and moreover we have $A_{0100}=
A_{0111}=0$, so that

\begin{equation} X_4=X_7=0. \end{equation}
The unitary relations (\ref{u3}) become
\begin{equation}\begin{array}{l}
 X_0^2+X_1^2+X_2^2+X_3^2+X_5^2+X_6^2=1,\medskip \\
 X_1\,X_6+X_2\,X_5=0. \end{array}\end{equation}
The six surviving $X_K$ can then be represented by four independent
parameters, $\lambda,\ \mu,\ \theta,\ \phi$, as follows:
\begin{equation}\begin{array}{lll}
  X_0=\sin\lambda\,\cos\mu, & \qquad & X_3=\sin\lambda\,\sin\mu,
  \medskip \\
  X_1=\cos\lambda\,\cos\theta\,\cos\phi, & \qquad &
  X_2=\cos\lambda\,\cos\theta\,\sin\phi, \medskip \\
  X_5=\cos\lambda\,\sin\theta\,\cos\phi, & \qquad &
  X_6=-\cos\lambda\,\sin\theta\,\sin\phi.\end{array}
\label{u4}\end{equation}

We are now ready to investigate the tradeoff between the information
acquired by Eve and the disturbance inflicted on the quantum system
that Bob receives. Let

\begin{equation} |\psi\rangle=\sum_m c_m\,|e_m\rangle, \end{equation}
be the pure state sent by Alice, e.g., one of the two signal states in
Eq.~(\ref{signals}). After Eve's intervention, the new state is

\begin{equation} |\psi'\rangle=\sum_{mn\beta}
c_m\,A_{mn\beta}\,|e_n,v_\beta\rangle, \label{psiprime}\end{equation}
and the density matrix of the combined system is $\rho'=
|\psi'\rangle\langle\psi'|$. (Here, we temporarily returned to using a
single Greek index $\beta$ for Eve's probe, instead of the composite
$rs$ index.) The reduced density matrices, for the two subsystems
considered separately, are then given by Eqs.~(\ref{rB}) and (\ref{rE}).
Explicitly, we have,

\begin{equation} (\rho_{\rm B}')_{mn}=\sum_\beta Y_{m\beta}\,Y_{n\beta},
\label{rBprime}\end{equation}

and
\begin{equation} (\rho_{\rm E}')_{\beta\gamma}=\sum_m
Y_{m\beta}\,Y_{m\gamma}, \label{rhoE} \end{equation}
where

\begin{equation} Y_{n\beta}=\sum_m c_m\,A_{mn\beta}.
 \label{Y} \end{equation}

The discrepancy rate observed by Bob is given by Eq.~(\ref{discrep}):

\begin{equation} D=1-\sum_{mn}c_m\,c_n\,(\rho_{\rm B}')_{mn}
  =1-\sum_\beta Z_\beta^2, \end{equation}
where

\begin{equation} Z_\beta=\sum_n c_n\,Y_{n\beta}=
\sum_{mn}c_m\,c_n\,A_{mn\beta}. \label{operation} \end{equation}
Explicitly, we have, when 01-symmetry holds,

\begin{equation} Z_{00}=c_0^2\,X_0+c_1^2\,X_3, \end{equation}
\begin{equation} Z_{01}=c_0^2\,X_1+c_0c_1\,(X_5+X_6)+c_1^2\,X_2,
 \end{equation}
\begin{equation} Z_{10}=c_0^2\,X_2+c_0c_1\,(X_5+X_6)+c_1^2\,X_1,
 \end{equation}
\begin{equation} Z_{11}=c_0^2\,X_3+c_1^2\,X_0. \end{equation}
With the help of Eqs.~(\ref{signals}), (\ref{S}), and (\ref{u4}), we
finally obtain
\begin{equation}\begin{array}{l}
 D=\cos^2\lambda\,\sin^2\theta
  -(S/2)\,\cos^2\lambda\,\sin 2\theta\,\cos 2\phi\medskip \\
  \qquad\qquad+\,(S^2/2)\,[\sin^2\lambda\;(1-\sin 2\mu)+
   \cos^2\lambda\,\cos 2\theta\;(1-\sin 2\phi)].\end{array}
  \label{D} \end{equation}

We now turn our attention to Eve, whose task is to gather
information about whether Alice sent $|0\rangle$ or $|1\rangle$. That
is, Eve must distinguish two different density matrices of type
(\ref{rhoE}), which differ by the interchange of $c_0$ and $c_1$. Let us
denote these density matrices as $\rho'_i$, with $i=0,1$.

Eve chooses a suitable POVM with elements $E_\mu$, as in
Eq.~(\ref{POVM}). From Eq.~(\ref{pmu}), the probability of getting
outcome $\mu$, following preparation $\rho'_i$, is

\begin{equation} P_{\mu i}={\rm Tr}\,(E_\mu\,\rho'_i). \end{equation}
Having found a particular $\mu$, Eve obtains the posterior probability
$Q_{i\mu}$ for preparation $\rho'_i$, by means of Bayes's rule [10]:

\begin{equation} Q_{i\mu}=P_{\mu i}\,p_i/q_\mu, \end{equation}
where

\begin{equation} q_\mu=\sum_j P_{\mu j}\,p_j, \end{equation}
is the prior probability for occurence of outcome $\mu$.

The Shannon entropy (Eve's level of ignorance), which initially was
$H=-\sum p_i\,\log p_i$, now is, after result $\mu$ was obtained,

\begin{equation} H_\mu=-\sum_i Q_{i\mu}\,\log Q_{i\mu}. \end{equation}
Therefore the mutual information (namely, Eve's average information
gain) is

\begin{equation} I=H-\sum_\mu q_\mu\,H_\mu. \end{equation}
This quantity depends both on the properties of Eve's probe (the various
$A_{mn\beta}$) and the choice of the POVM elements $E_\mu$.\\[10mm]

\begin{center}{\bf III. OPTIMIZATION} \end{center}\bigskip

If Eve wants to maximize the mutual information $I$, she has to choose the
POVM elements $E_\mu$ in an optimal way. This is a complicated nonlinear
optimization problem, for which there is no immediate solution. There
are, however, useful theorems, due to Davies [11]. Firstly, an optimal
POVM consists of matrices of rank one:

\begin{equation} E_\mu=|w_\mu\rangle\langle w_\mu|. \end{equation}
(To be precise, there may be POVMs made of matrices of higher rank, that
give the same mutual information as these optimal matrices of rank one,
but they can never give more mutual information.)

Secondly, the required number, $N_w$, of different vectors
$|w_\mu\rangle$, is bracketed by

\begin{equation} N\leq N_w\leq N^2, \label{Davies} \end{equation}
where $N$ is the dimensionality of Hilbert space. A rigorous proof of
this relationship, given by Davies [11], is fairly intricate. A
plausibility argument (not a real proof) can be based on the reasoning
subsequent to Eq.~(\ref{Phimu}). If the POVM is implemented by an
instrument obeying the laws of quantum mechanics, the interaction with
this instrument is unitary. Therefore, the instrument's final state
after the interaction must reside in a fixed subspace of no more than
$N^2$ dimensions (if the initial state of the instrument was pure). When
we perform a von \mbox{Neumann} measurement on the instrument---which is
the upshot of the POVM procedure---it is thus plausible that it should
never be necessary to involve more than $N^2$ distinct outcomes.

In the present case, $N=4$ (the number of dimensions of Eve's probe),
and Davies's theorem guarantees that Eve does not need more than 16
different vectors $|w_\mu\rangle$, subject to the constraint
$\sum_\mu|w_\mu\rangle\langle w_\mu|=\openone$.  Moreover, while there
are cases where the upper limit in (\ref{Davies}) is indeed reached (an
example is given in Davies's work), there also are cases for which it is
known that $N_w$ need not exceed $N$. It is so when we have to
distinguish two pure states, or even two density matrices of rank~2,
lying in the same two-dimensional subspace of Hilbert space [12]. It has
been conjectured [12] that this is also true for any two density
matrices of arbitrary rank. In the absence of a formal proof, we tested
that conjecture numerically, for more than a hundred pairs of randomly
chosen density matrices, with $N=3$ or $N=4$.  Using the Powell
algorithm [13], we tried various values of $N_w$ in the range given by
Eq.~(\ref{Davies}). In all these tests, it never happened that the
number of vectors had to exceed $N$ (namely, whenever we tried $N_w>N$,
we found that some of the optimized vectors were parallel, and there
were only $N$ independent $|w_\mu\rangle$.)

Therefore, in the present case, we assume Eve only has to find the
optimal 4-dimensional {\it ortho\-normal\/} basis $\{|w_\mu\rangle\}$.
(This result might have been expected, in view of the above argument for
the plausibility of Eq.~(\ref{Davies}), because in an optimal unitary
evolution there cannot be more than four different final outputs, if there
are two inputs.)  We thus have now a standard optimization problem, which
can be solved numerically (the orthonormality constraint must be handled
carefully, though, so that iterations converge). However, Eve has an
additional problem, which is to find the optimal unitary interaction for
her probe, in Eq.~(\ref{u0}). She must therefore include, in the
optimization procedure, the four angles $\lambda,\ \mu,\ \theta,\ \phi$,
defined in Eq.~(\ref{u4}). Moreover, she may also want to control the
disturbance $D$, given by Eq.~(\ref{D}).

Many different tradeoffs can be chosen, when we want to maximize $I$ and
to minimize $D$. A simple figure of merit could be $M=I-kD$, where the
positive coefficient $k$ expresses the value of the information $I$,
compared to the cost of causing a disturbance $D$.  We could also
imagine other, more complex figures of merit, involving nonlinear
functions of $I$ and $D$. With crypto\-graphical applications in mind,
we investigated the problem of maximizing $I$ subject to the constraint
$D\leq D_{\rm tol}$, so that the disturbance be less than a certain
tolerable one. This was done by maximizing the function
$M=I-1000\,(D-D_{\rm tol})^2$, for many randomly chosen values of
$\alpha$ in Eq.~(\ref{signals}).

In all the cases that we tested, the optimization procedure led to
$\lambda=0$ (or to an integral multiple of $\pi$) in Eq.~(\ref{u4}).
This implies $X_0=X_3=0$, and since we already have $X_4=X_7=0$, this
means that $\forall rs,\ A_{rs00}=A_{rs11}=0$, and therefore
$|\Phi_{00}\rangle=|\Phi_{11}\rangle=0$. We remain with only
$|\Phi_{01}\rangle$ and $|\Phi_{10}\rangle$. In other words, Eve's
optimal probe has only two dimensions, not four.

We have no formal proof for this result, which was found by numerical
experiments. However, this result is quite plausible: it is clear from
Eq.~(\ref{D}) that $D$ is an even function of $\lambda$, and therefore
is extremized when $\lambda=0$. Unfortunately, it is more difficult to
evaluate explicitly the mutual information $I$, which is a complicated
function of the matrix elements $(\rho_{\rm E}')_{mn,rs}$ in
Eq.~(\ref{rhoE}). However, when we write explicitly these matrix
elements, we see that they are even or odd functions of $\lambda$,
according to the parity of the sum of indices, $(m+n+r+s)$. This
symmetry property then holds for any product of such matrices, and the
trace of any such product always is an even function of $\lambda$. Since
$I$ is a scalar, {\it i.e.,\/} is invariant under a change of the basis,
it is plausible that $I$ can be written, or at least approximated, by
expressions involving only these traces, so that $I$ also is an even
function of $\lambda$. Therefore $\lambda=0$ is an extremum of our
figure of merit, and it might be possible to prove, with some effort,
that $\lambda=0$ indeed gives the global maximum of the figure of merit.
Anyway, the validity of this result is likely to be restricted to the
highly symmetric case where Alice prepares two equiprobable pure states,
as in Eq.~(\ref{signals}).

However, once this result is taken for granted, the calculation becomes
considerably simpler, and can be done analytically, rather than
numerically. First, we note that, by virtue of the 01-symmetry, Eve's
two density matrices can be written as

\begin{equation} \rho'_0
 =\left(\matrix{a & c \cr c & b}\right),\qquad{\rm and}\qquad  \rho'_1
 =\left(\matrix{b & c \cr c & a}\right),\label{rhoprime}\end{equation}
with $a+b=1$. These two matrices have the same determinant,

\begin{equation} d=ab-c^2\geq 0. \end{equation}
In that case, the mutual information that can be extracted from them is
explicitly given by [12, 14]:

\begin{equation} I=[(1+z)\,\log(1+z)+(1-z)\,\log(1-z)]/2,
 \label{I} \end{equation}
where

\begin{equation} z=[1-2d-{\rm Tr}\,(\rho'_0\;\rho'_1)]^{1/2}
 =(1-4ab)^{1/2}. \label{z} \end{equation}
We therefore need only the diagonal elements in (\ref{rhoE}). These are,
by virtue of (\ref{u4}) and (\ref{Y}),
\begin{eqnarray}
(\rho_{\rm E}')_{01,01} & = & \sum_n Y_{n01}^2, \\
 & = & (c_0\,X_1+c_1\,X_6)^2+(c_0\,X_5+c_1\,X_2)^2, \\
 & = & (1+\cos 2\alpha\,\cos 2\phi)/2, \end{eqnarray}
and likewise

\begin{equation} (\rho_{\rm E}')_{10,10}=(1-\cos 2\alpha\,\cos 2\phi)/2,
\end{equation}
where $\alpha$ is the angle defined in Eq.~(\ref{signals}). (Here, to
conform with our earlier notations, each one of the two dimensions of
the probe's space is denoted by a double index, 01 or 10.)  We thus
obtain

\begin{equation} z=[1-4\,(\rho_{\rm E}')_{01,01}\,
 (\rho_{\rm E}')_{10,10}]^{1/2}=\cos 2\alpha\,\cos 2\phi. \end{equation}
When substituted in Eq.~(\ref{I}), this result gives a remarkably simple
expression for the mutual information.  In particular, $I$ does not
depend on $\theta$.

The discrepancy rate $D$, given by Eq.~(\ref{D}), also simplifies:

\begin{equation} D=\sin^2\theta-(S/2)\,\sin2\theta\,\cos2\phi+
 (S^2/2)\,\cos2\theta\,(1-\sin2\phi), \label{D1} \end{equation}
whence
\begin{equation}
 2D =  1-S\,\cos2\phi\,\sin2\theta-
       [1-S^2\,(1-\sin2\phi)]\,\cos2\theta.
\end{equation}
For each $\phi$, the angle $\theta=\theta_0$ making $D$ minimal is given
by

\begin{equation} \tan2\theta_0=S\,\cos2\phi\,/\,[1-S^2\,(1-\sin2\phi)],
 \label{Dmin} \end{equation}
and that minimal value of $D$ is

\begin{equation}
 2D_0= 1- \{S^2\,\cos^2 2\phi+[1-S^2\,(1-\sin2\phi)]^2\}^{1/2}.
 \label{MinDist} \end{equation}

Let us consider various values of $\phi$. For $\phi=0$, we obtain the
maximal value of $I$:

\begin{equation} I_{\rm max}=\log2+\cos^2\alpha\,\log(\cos^2\alpha)
   +\sin^2\alpha\,\log(\sin^2\alpha), \label{Imax} \end{equation}
as could have been found more directly. The minimal disturbance
corresponding to this $I_{{\rm max}}$ is

\begin{equation} D_1=[1-(1-S^2+S^4)^{1/2}]/2. \label{Fid} \end{equation}

Clearly, it is possible to have $D_0<D_1$ only
by accepting $I<I_{{\rm max}}$.  By solving Eq.~(\ref{MinDist}) for
$\phi$ and using Eq.~(\ref{Fid}), one gets an explicit relation
between the maximal information and the minimal disturbance caused by
the measurement.  This is given by Eq.~(\ref{I}) with

\begin{equation} z=\cos2\alpha\left[1-\left(1-\sqrt{
D_0(1-D_0)/D_1(1-D_1)}\,
\right)^{\!\! 2}\,\,\right]^{1/2}.  \end{equation}
This relation completely specifies the information-disturbance tradeoff.
The result is plotted in Fig.~3 for three values of the angle $\alpha$
defined by Eq.~(\ref{signals}), namely $\alpha=\pi/16,\ \pi/8$, and
$\pi/5$ (these are the values that were investigated in ref.~[5]).

The limit $D_0\to0$ is obtained for $\phi=(\pi/4)-(\epsilon/2)$, with
$\epsilon\to0$. We then have

\begin{equation} I\to z^2/2\simeq(\epsilon\,\cos2\alpha)^2/2,
 \end{equation}
and

\begin{equation}
D_0\to\epsilon^4\,(S^2-S^4)/16\simeq(I\,\tan2\alpha)^2/4. \end{equation}
The quadratic behavior, $D_0\sim I^2$, which was derived for the pair of
non-orthogonal signals in Eq.~(\ref{signals}), may however not hold for
more complicated types of quantum information [15], such as the two
orthogonal pairs in ref.~[1].

Finally, let us examine the correlation between the result observed by
Eve and the quantum state delivered to Bob. We can write
Eq.~(\ref{psiprime}) as

\begin{equation} |\psi'\rangle=
 \sum_{\beta} |\psi'_\beta\rangle\otimes|v_\beta\rangle,
 \label{EPR} \end{equation}
where

\begin{equation} |\psi'_\beta\rangle=
 \sum_{mn} c_m\,A_{mn\beta}\,|e_n\rangle, \end{equation}
is (except for normalization) the state received by Bob whenever
Eve observes outcome $\beta$. For example, if Alice sends $|0\rangle$
and Eve observes $|v_{01}\rangle$, Bob receives
\begin{eqnarray} |\psi'_{01}\rangle & = &
(c_0\,X_1+c_1\,X_6)\,|e_0\rangle+(c_0\,X_5+c_1\,X_2)\,|e_1\rangle,\\
 & = & (\cos\alpha\,\cos\theta\,\cos\phi-
\sin\alpha\,\sin\theta\,\sin\phi)\,|e_0\rangle \nonumber\\
 & & \qquad +\,(\cos\alpha\,\sin\theta\,\cos\phi+
\sin\alpha\,\cos\theta\,\sin\phi)\,|e_1\rangle.\end{eqnarray}
Note that

\begin{equation} \|\psi'_{01}\|^2=
  \cos^2\alpha\,\cos^2\phi+\sin^2\alpha\,\sin^2\phi, \end{equation}
is the probability that Bob gets $|\psi'_{01}\rangle$ when Alice sends
$|0\rangle$ and Eve observes $|v_{01}\rangle$.

Let us consider two extreme cases. If $\phi=\pi/4$, so that Eve obtains
no information, we may choose $\theta=0$ in accordance with
Eq.~(\ref{Dmin}), and it then follows from Eq.~(\ref{D1}) that there is
no disturbance at all. Indeed, in that case, the $U$ matrix in
Eq.~(\ref{u0}) simply is a unit matrix.

On the other hand, if $\phi=0$ so that Eve acquires all the accessible
information, Bob receives, with probability $\cos^2\alpha$, a state

\begin{equation}
|\tilde\psi'_{01}\rangle=\cos\theta\,|e_0\rangle+\sin\theta\,|e_1\rangle.
  \label{0prime} \end{equation}
(The tilde indicates that this state has been normalized.) The angle
$\theta$ that minimizes $D$ is given by Eq.~(\ref{Dmin}), which now
becomes

\begin{equation} \tan2\theta=S/(1-S^2)=\sin2\alpha/\cos^22\alpha=
  \tan2\alpha/\cos2\alpha. \label{alphaprime} \end{equation}
Everything happens as if, when Eve observes the state closest to
$|0\rangle$, she sends to Bob, not $|0\rangle$, but a slightly {\it
different\/} state $|0'\rangle$, with a new angle $\theta$, slightly
larger than $\alpha$. For example, if $\alpha=22.5^\circ$, we have
$\theta=27.3678^\circ$. These angles are illustrated in Fig.~4. It must
however be pointed out that, in the scenario described in Fig.~1, Eve
releases Bob's particle {\it before\/} observing her probe. What she
actually has to do is to make them interact with the appropriate $U$,
and this guarantees that the final state is correctly correlated, as in
Eq.~(\ref{EPR}).\\[10mm]

\begin{center}{\bf IV. OTHER TRADEOFF CRITERIA}\end{center}\bigskip

Until now, we used $D$ in Eq.~(\ref{discrep}) as a measure of the
disturbance: this was the probability for an observer to find the
quantum system in a state orthogonal to the one prepared by Alice. This
may not always be the most useful criterion, and in some cases it indeed
is a very poor one. For example, if the two states in
Eq.~(\ref{signals}) have $\alpha$ close to $\pi/4$, the states sent to
Bob will be even closer to $\pi/4$, as may be seen from
Eq.~(\ref{alphaprime}). The states themselves change very little, but
the information that they carry is drastically reduced, as the following
example shows.

Consider the case $\alpha=\pi/5$ (depicted by the lowest line in
Fig.~3). Eve then has $I_{\rm max}=0.048536$. Let us rename this
expression $I_{\rm AE}$ (the mutual information for Alice and Eve).
Two different mutual informations can likewise be defined for Bob:
$I_{\rm EB}$, namely, what Bob may be able to know on the result
registered by Eve, and $I_{\rm AB}$, what he may still be able to know
on the original state, prepared by Alice.

The calculation of $I_{\rm EB}$ is easy. As explained after
Eq.~(\ref{alphaprime}), everything happens as if Eve would knowingly
send to Bob one of two pure states, like those in Eq.~(\ref{signals}),
but with $\alpha=36^\circ$ replaced by $\theta=42.1332^\circ$. We then
have, from Eq.~(\ref{Imax}), $I_{\rm EB}=0.0049987$, about one tenth of
$I_{\rm AE}$.

What Bob may still be able to know about the state that was sent by
Alice is even less than that. Bob receives the quantum system in a state
described by the density matrices (\ref{rBprime}). Due to 01-symmetry,
these matrices have the same form (\ref{rhoprime}) as those of Eve, and
the mutual information $I_{\rm AB}$ is again given by Eqs.~(\ref{I}) and
(\ref{z}). Now, however,

\begin{equation} a=(\rho_{\rm B}')_{00}=Y_{0,01}^{\:2}+Y_{0,10}^{\:2}
 =(c_0\,X_1+c_1\,X_6)^2+(c_0\,X_2+c_1\,X_5)^2,\end{equation}
and
\begin{equation} b=(\rho_{\rm B}')_{11}=Y_{1,01}^{\:2}+Y_{1,10}^{\:2}
 =(c_0\,X_5+c_1\,X_2)^2+(c_0\,X_6+c_1\,X_1)^2.\end{equation}
It follows that, regardless of the value of $\phi$

\begin{equation} z=\cos2\alpha\,\cos2\theta. \end{equation}
If we now take $\theta$ given by Eq.~(\ref{alphaprime}), we obtain $z=
0.0308718$, whence $I_{\rm AB}=0.0004766$. This is more than a hundred
times smaller than the mutual information Bob could have had if Eve's
probe had not been in the way!  Thus, in that sense, Eve caused a major
disturbance, even though it was as small as it could be by the previous
criterion (for the given amount of information she gains).

Note however that Eve, who controls both $\phi$ and $\theta$, could just
as well set $\theta=0$. In that case, Bob would be able to recoup all
the mutual information sent by Alice, simply by measuring the orthogonal
states forwarded on to him.  Nevertheless this scenario can hardly count
as a minimally disturbing intervention on Eve's part, because in that
case $D=S^2/2=0.452254$, as can be seen from Eq.~(\ref{D1}).

What appears to be needed is a measure of disturbance that is itself of
an information theoretic nature. There are many ways of comparing the
states sent by Alice to the states received by Bob, that have more
information-theory flavor than the measure used in the previous
sections. For instance, one might consider using the Kullback-Leibler
relative information~[16]. The latter quantifies the discrepancy
between the frequencies of outcomes for a quantum measurement on Alice's
states, versus that same measurement on Bob's states~[17].  Or, one
might consider using the Chernoff information~[16], which quantifies
Bob's difficulty in guessing whether Eve has tampered with the state (in
a given way) or not~[17].  In any case, the best measure of disturbance
is the one that is relevant to the actual application in which we are
interested.\\[10mm]

\begin{center}{\bf ACKNOWLEDGMENTS}\end{center}\bigskip

CAF thanks H.~Barnum and D.~Mayers for discussions. Part of this work
was done during the ``Quantum Computation 1995'' workshop held at the
Institute for Scientific Interchange (Turin, Italy) and sponsored by
ELSAG-Bailey.  Work by CAF was supported in part by the Office of Naval
Research (Grant No.~N00014-93-1-0116). Work by AP was supported by the
Gerard Swope Fund and the Fund for Encouragement of Research.\\[10mm]

\newpage\frenchspacing
\begin{enumerate}
\item  C. H. Bennett and G. Brassard, in {\em Proceedings of IEEE
 International Conference on Computers, Systems and Signal Processing,
 Bangalore, India} (IEEE, New York, 1984) p.~175.
\item A. K. Ekert, Phys. Rev. Lett. {\bf 67}, 661 (1991).
\item C. H. Bennett, Phys. Rev. Lett. {\bf 68}, 3121 (1992).
\item W. G. Unruh, Phys. Rev. D {\bf 18}, 1764 (1978); {\bf 19}, 2888
(1979).
\item A. K. Ekert, B. Huttner, G. M. Palma, and A. Peres, Phys. Rev. A
{\bf 50}, 1047 (1994).
\item  C. H. Bennett, F. Bessette, G. Brassard, L. Salvail, and J.
Smolin, J. Crypto. {\bf 5}, 3 (1992).
\item J. von Neumann, {\it Mathematische Grundlagen der
Quantenmechanik\/} (Springer, Ber\-lin, 1932) [transl. by E.~T.~Beyer:
{\it Mathematical Foundations of Quantum Mechanics\/} (Princeton Univ.
Press, Princeton, 1955)] Chapt. 3.
\item C. W. Helstrom, {\it Quantum Detection and Estimation Theory\/}
(Academic Press, New York, 1976) pp.~80--84.
\item A. Peres, {\it Quantum Theory: Concepts and Methods\/} (Kluwer,
 Dordrecht, 1993) pp. 282--288.
\item W. Feller, {\it An Introduction to Probability Theory and its
Applications\/} (Wiley, New York, 1968) vol.~I, p.~124.
\item E. B. Davies, IEEE Trans. Inform. Theory {\bf IT-24}, 596 (1978).
\item L. B. Levitin, in {\it Quantum Communication and Measurement\/},
 ed. by V.~P. Belavkin, O. Hirota, and R. Hudson (Plenum Press, New
 York, 1995) p.~439.
\item W. H. Press, S. A. Teukolsky, W. T. Vetterling, and B. P.
Flannery, {\it Numerical Recipes\/} (Cambridge University Press,
Cambridge, 1992) Chapt. 10.
\item C. A. Fuchs and C. M. Caves, Phys. Rev. Lett. {\bf 73}, 3047
 (1994).
\item T. Mor (private communication).
\item T. M. Cover and J. A. Thomas, {\it Elements of Information Theory\/}
(Wiley, New York, 1991) Chapt. 12.
\item C. A. Fuchs, {\it Distinguishability and Accessible Information
in Quantum Theory}, Ph.~D. Thesis, University of New Mexico (1995).
\end{enumerate}
\nonfrenchspacing

\vfill
\parindent 0mm
{\bf Captions of figures}\bigskip

FIG. 1. \ Eve's probe interacts unitarily (U) with the particle sent by
Alice to Bob, and is then subjected to a generalized measurement
(M).\bigskip

FIG. 2. \ Choice of basis for signal states (a) and probe's states
(b).\bigskip

FIG. 3. \ Maximal mutual information $I$ obtainable for a given
disturbance $D$, for two equiprobable pure input signals. The angle
$\alpha$ is defined by Eq.~(\ref{signals}). The dashed lines represent
the maximal obtainable $I$, which cannot be exceeded by accepting a
further increase of $D$.\bigskip

FIG. 4. \ States $|0\rangle$ and $|1\rangle$ are sent by Alice, and
states $|0'\rangle$ and $|1'\rangle$ are resent by Eve, so as to cause
the least possible disturbance rate in Bob's observations.

\end{document}